\documentclass[a4paper]{jpconf}
\usepackage{graphicx}

\usepackage{graphicx}
\usepackage{dcolumn}
\usepackage{bm}
\usepackage{verbatim}
\usepackage{epsfig}
\usepackage{epstopdf}

\newcommand\beq{\begin{equation}}
\newcommand\eeq{\end{equation}}
\newcommand\bea{\begin{eqnarray}}
\newcommand\eea{\end{eqnarray}}

\def\half{\frac {1} {2}}

\def\x0{{{\bf x}_0}}

\def\implies{\Rightarrow }

\def \MR {\rm MR}
\def\LG{\rm LG}
\def\NSIT{\rm NSIT}
\def\NIM{\rm NIM}
\def\Ind{\rm AoT}
\def\Ind{\rm Ind}

\begin{document}


\title{Necessary and sufficient conditions for macrorealism using two and three-time Leggett-Garg inequalities}

\author{J.J.Halliwell}%

\address{Blackett Laboratory, Imperial College, London SW7 2BZ, UK }

\ead{j.halliwell@imperial.ac.uk}



\begin{abstract}
The Leggett-Garg (LG) inequalities were introduced, as a temporal parallel of the Bell inequalities, to test macroscopic realism -- the view that a macroscopic system evolving in time possesses definite properties which can be determined without disturbing the future or past state.
The original LG inequalities are only a necessary condition for macrorealism, and are therefore not a decisive test. We argue, for the case of measurements of a single dichotomic variable $Q$,
that when the original four three-time LG inequalities are augmented with a set of twelve two-time inequalities also of the LG form, Fine's theorem applies and these augmented conditions are then both necessary and sufficient. 
A comparison is carried out with the alternative necessary and sufficient conditions for macrorealism 
based on no-signaling in time conditions which ensure that all probabilities for $Q$ at one and two times are independent of whether earlier or intermediate measurements are made.
We argue that the two tests differ in their implementation of the key requirement of non-invasive measurability so are testing different notions of macrorealism, and these notions are elucidated.
\end{abstract}

\section{Introduction}

The purpose of this paper is to develop, discuss and compare various conditions that have been proposed to test macrorealism, the idea that a time-evolving macroscopic system can possess definite properties at a number of times uninfluenced by measurements of it. The aim in particular is to present conditions for macrorealism which are necessary and sufficient, and also to elucidate the different, inequivalent, notions of non-invasiveness involved in such tests.
This paper is in part a more streamlined statement of a formulation given in Refs.\cite{HalQ,Hal0} and removes some inessential assumptions contained therein.

\subsection{Macrorealism and the Leggett-Garg inequalities}

Macrorealism (MR) was proposed by way of analogy to the notion of local realism for spatially entangled systems and leads to a set of inequalities obeyed by the temporal correlation functions of a single system, the Leggett-Garg inequalities  \cite{LG1,L1,ELN}.
Most investigations to date focus on a single dichomotic variable $Q$ which is measured in various ways at three (or more) times leading to the determination of the temporal correlation functions of the form,
\beq
C_{12} = \langle Q(t_1) Q(t_2) \rangle.
\label{corr}
\eeq
These are argued, for a macrorealistic theory, to obey the Leggett-Garg (LG) inequalities,
\bea
1 + C_{12} + C_{23} + C_{13} & \ge & 0,
\label{LG1}
\\
1 - C_{12} - C_{23} + C_{13} & \ge & 0,
\label{LG2*}
\\
1 + C_{12} - C_{23} - C_{13} & \ge & 0,
\\
1 - C_{12} + C_{23} - C_{13} & \ge & 0,
\label{LG4}
\eea
which are identical in mathematical form to the Bell inequalities. Measurements at four times lead to a set of eight LG inequalities identical in mathematical form to the CHSH inequalities.

To derive these inequalities, the notion of macrorealism is broken down into three separate assumptions. These are:

\noindent{\bf 1.} Macrorealism {\it per se} (MRps): the system is in one of the states available to it at each moment of time.

\noindent{\bf 2.} Non-invasive measurability (NIM): it is possible in principle to determine the state of the system without disturbing the subsequent dynamics.

\noindent{\bf 3.} Induction (Ind): future measurements cannot affect the present state. 

These assumptions ensure the existence of an underlying joint probability distribution from which the LG inequalities readily follow.
Any experimental test thus tests the combination of these assumptions. Induction is always taken for granted so what is being tested is the combination of MRps and NIM. To ensure NIM, Leggett and Garg proposed that the measurement of the correlation functions be carried out using ideal negative measurements, in which the detector is coupled to, say, only the $Q=+1$ state, at the first time, and a null result then permits us to deduce that the system is in the $Q=-1$ state but without any interaction taking place, from the macrorealistic perspective. This procedure rules out alternative classical explanations of the correlation functions \cite{Mon} analogous to the way in which signaling is ruled out in Bell experiments and has been successfully implemented in a number of recent experiments \cite {Knee,Rob,KBLL,EmINRM}. Many other experimental tests of the LG inequalities have also been carried out, on a variety of different physical systems (see the extensive list of references in Refs.\cite{Hal0,ELN}).

Maroney and Timpson \cite{MaTi} have argued that MRps actually comes in three different varieties, only one of which can be ruled out by LG test, namely
``operational eigenstate mixture macrorealism'', which includes spontaneous collapse models, such as those of the GRW type \cite{GRW}. Almost all LG tests are focused on this variety of MRps and we do so here.
Of the other two varieties of MRps,
the most significant one is realist theories of the Bohmian type, which would need some sort of locality arguments in order to be ruled out \cite{MaTi,Bac}.
The remaining type have the form of restricted Bohmian theories, and include the Kochen-Specker model \cite{KS} for two-dimensional Hilbert spaces, but for Hilbert spaces of dimension greater than $3$ these can be ruled out \cite{Ma2,Ma3}.

\subsection{Different interpretations of NIM}

In the various papers on tests of macrorealism, NIM appears to 
be interpreted in a number of different ways depending on exactly how the measurements are carried out.
Most experimental tests of the LG inequalities measure the three correlation functions in three different experiments, analogous to the Bell case, with each involving measurements at just two times and non-invasiveness is required only in each separate experiment. Furthermore, even within each experiment involving two times, there are different choices, as we shall see in more detail. However, some approaches assume
a stronger reading of NIM -- that it should not make any difference if one, two or three sequential measurements are made in the same experiment.
Both of these versions of NIM permit access to the information required to determine whether MRps holds, but since MR is defined to be the conjunction of both MRps and NIM, it means that there are {\it different versions} of MR depending on how the NIM requirement is implemented. These differences will be important for what follows.

For convenience we will denote the stronger version of NIM, entailing non-invasiveness for sequential measurements at three times, by $\NIM_{seq}$, and the weaker one, in which NIM is only satisfied in a piecewise (pw) way in each one of a number of  experiments, by $\NIM_{pw}$. These different characterizations will be further refined as required.

\subsection{Parallels with Bell experiments}

The LG framework for testing macrorealism was designed by analogy with Bell experiments and 
it does indeed have some genuinely close parallels (reviewed in Ref.\cite{Hal0}).
However, the analogy fails at a number of points. As Maroney and Timpson have argued \cite{MaTi}, Bell and LG tests are not methodologically on a par since the notion of non-invasiveness typically carries some model-dependent assumptions so is difficult to motivate as a general feature, unlike local causality in Bell experiments.

This general point relates to our main question of interest, which is that of sufficient conditions for macrorealism. 
The LG inequalities are necessary conditions but they are not sufficient, as has been noted by a number of authors \cite{KoBr,Cle}.
By contrast in Bell experiments, Fine's theorem \cite{Fine,HalFine}
guarantees that the Bell \cite{Bell} or CHSH \cite{CHSH} inequalities are both necessary and sufficient conditions for the existence of an underlying probability matching the given correlation functions, and so are necessary and sufficient conditions for local realism.  This means that the Bell or CHSH inequalities are a decisive test.

The point at which Fine's theorem fails to apply to the LG framework relates to the description of the system at two moments of time. Suppose we carry out sequential measurements of $Q$ at $t_1$ and $t_2$, yielding probability $p_{12} (s_1,s_2)$ and in a separate experiment, measure $Q$ at $t_2$ only, yielding probability $p_2 (s_2)$. For Fine's theorem to hold, it would be necessary for conditions of the form
\beq
\sum_{s_1} p_{12} (s_1, s_2) = p_2 (s_2),
\label{NSIT}
\eeq
to hold, since such conditions ensure that the pairwise probabilities such as  $p_{12} (s_1,s_2)$, $p_{23} (s_2,s_3)$ etc are compatible with each other. Conditions of the form Eq.(\ref{NSIT}) are referred to as ``no signaling in time'' (NSIT) conditions \cite{KoBr,Cle}. They were proposed as a clear analogy to no-signaling in Bell experiments but are not satisfied in general (except for special parameter choices).

However, this difficulty clearly relates to the particular way in which NIM is implemented. Eq.(\ref{NSIT}) is naturally linked to $\NIM_{seq}$, but one could also contemplate a less direct measurement scheme for the two-time probabilities, consistent with $\NIM_{pw}$, which yields a different two-time probability compatible with Eq.(\ref{NSIT}).
From a quantum-mechanical perspective this corresponds to the fact that probabilities for non-commuting quantities can often be defined but they are not unique, with different measurements procedures giving different candidate probabilities, and furthermore, their interpretation as probabilities comes with restrictions.

\subsection{Necessary and sufficient conditions for macrorealism}

We now outline two different ways in which necessary and sufficient conditions for MR can be defined, making use of the two different versions of NIM outlined above.
The first way
is to work with the weaker form of non-invasiveness, $\NIM_{pw}$, and
stay as close as possible to the original LG framework, in which the three correlation functions are determined in a number of different runs. To sidestep the failure of Eq.(\ref{NSIT})
we seek another way of determining a candidate probability for the system at two times which does not involve sequential measurements. 
Our approach is to use a number of different non-invasive experiments to determine the moments of the system at two times and then use a set of two-time LG inequalities to determine indirectly whether or not
a probability distribution at two times exsists.
This approach yields a set of two-time and three-time LG inequalities which, when the two-time LG inequalities hold, have a mathematical form identical to that of the Bell system and are therefore necessary and sufficient conditions for MR. Hence, the desired parallel with the Bell system and a decisive test for MR is achieved using an augmented set of LG inequalities measured in a judiciously chosen set of runs.  The measurements are non-invasive, by design, so this protocol is, under ideal conditions, most accurately thought of as a direct test of MRps.

The second way is to follow the much stronger reading of NIM outlined above, $\NIM_{seq}$, and restrict to initial states and other parameter ranges so that
relationships of the form Eq.(\ref{NSIT}) hold for sequential measurements. 
In particular, Clemente and Kofler \cite{Cle} proposed a scheme in which the underlying three-time probability $p_{123}(s_1,s_2,s_3) $
is determined in a single experiment by sequential measurements at all three times, subject to a set of two- and three-time NSIT conditions, similar to Eq.(\ref{NSIT}). We review this work.
When these conditions hold, the probability $p_{123}(s_1,s_2,s_3)$ is a properly defined probability for a set of three independent variables, and
hence the set of NSIT conditions are a necessary and sufficient condition for MR.
The LG inequalities are not involved in this sort of test, but are clearly implied by the set of NSIT conditions. These conditions test a combination of $\NIM_{seq}$ and MRps (and induction).
From a quantum-mechanical point of view, it is very different to a Bell test since it involves sequential measurement of incompatible quantities, but 
this test is of the same type as some of the ``coherence witness'' tests proposed recently \cite{Rob,Wit,Ema}.

These two possibilities clearly delineate two extremes in terms of how strongly or weakly NIM is implemented. We will also find intermediate possibilities that involve combinations of both. (We also note here a possible connnection with the so-called Wigner Leggett-Garg inequalities, which lie midway between the LG inequalities and no-signaling conditions \cite{WLG}).

Note also that in talking about measurements which we refer to as ``non-invasive'', we have in mind a theoretical ideal situation. 
In practice experimental clumsiness is difficult to eliminate and this leaves loopholes for alternative explanations of the results \cite{deco}. See Refs.\cite{Knee,Rob} for further discussions of how this may be handled in specific experiments.

Tests of macrorealism involving the augmented LG inequalities are described in Section 2, and tests involving NSIT conditions are described in Section 3. Some quantum-mechanical aspects of the LG and NSIT approaches are briefly discussed in Section 4 along with a simple property of coherence witnesses.
We summarize and conclude in Section 5.

\section{Necessary and sufficient conditions for macroealism using Leggett-Garg inequalities}

We now exhibit a set of necessary and sufficient conditions for macrorealism using an augmented set of Leggett-Garg inequalities and using $\NIM_{pw}$.
We suppose that we can carry out a set of experiments to determine the averages and second order correlation functions of $Q(t)$ at three times. $\NIM_{pw}$ is the requirement that all quantities are measured non-invasively, but the measurements need be non-invasive only within each experiment -- there is no requirement that non-invasiveness should persist if different experiments are combined. This is a natural parallel with the Bell case \cite{Hal0}. The clearest way to accomplish this  is to do a total of six different experiments, in which 
the three correlation functions $C_{12}$, $C_{23}$, $C_{13}$ are determined in three different experiments, as in traditional LG tests, using a suitable non-invasive measurement protocol;
and
$\langle Q_1 \rangle$, $ \langle Q_2 \rangle$, $ \langle Q_3 \rangle $ are determined in three further experiments, in which there is clearly no issue of invasiveness since only a single measurement is made in each case. (We use the convenient notation $ Q_i  = Q(t_i)$ for $i=1,2,3$.)

Non-invasive protocols exist in which the correlation function {\it and nothing else} is measured \cite{HalLG2,HalLG3}, which is why it is reasonable to suppose that $\langle Q_1 \rangle$ and $C_{12}$ are determined in genuinely seperate experiments. 
Having said that, when we measure correlation functions such as $C_{12}$ with ideal negative measurements (as suggested in the original LG framework), this usually entails a single experiment in which the full two-time probability $p(s_1,s_2)$ is measured,
from which one could also read off $\langle Q_1 \rangle$. Hence it may be possible to reduce the six different experiments to three, as long as there is a clear argument for non-invasiveness.
However, for logical clarity, we make no assumptions about which pairs of moments can be measured in the same experiment (in contrast to Ref.\cite{Hal0} which invokes quantum mechanics to argue that certain pairs of moments are compatible).

Macrorealism is the question as to whether or not there exists an underlying probability $p(s_1,s_2,s_3)$ matching the six moments measured in the way described above. We build this up in three steps.

First, the single time probabilties are given by
\beq
p(s_i) = \frac{1}{2} \left( 1 + s_i \langle Q_i \rangle \right),
\eeq
for $i=1,2,3$, and these are non-negative by construction. Second, there are three two-time probabilties, given by
\beq
p(s_i,s_j) = \frac{1}{4} \left( 1 + s_i \langle Q_i \rangle + s_j \langle Q_j \rangle + s_i s_j C_{ij} \right),
\label{qmom}
\eeq
where $ij =12,13,23$. (See Refs.\cite{HaYe,Kly} for more information on these useful moment expansions).
Because the averages $ \langle Q_i \rangle$ are measured in such a way that there can be no disturbance from an earlier measurement,
these two-time probabilities automatically obey all compatibility conditions of the form,
\beq
\sum_{s_i} p(s_i,s_j) = p(s_j) = \sum_{s_k} p(s_j,s_k).
\label{comp}
\eeq
These relations are of course mathematically identical to the NSIT conditions, Eq.(\ref{NSIT}), but there is no sense in which they indicate the absence of ``signaling'', since the two-time probabilities are assembled indirectly from different experiments, not measured sequentially in a single experiment.
Instead these relations are simply the compatibility relations between the two-time probabilities that are required for Fine's theorem to apply.

In a macrorealist theory in which the averages and correlation function are non-invasively measured, the two-time probabilities Eq.(\ref{qmom}) are guaranteed to be non-negative. This follows very easily from a simple argument similar to the derivation of the LG inequalities: we have
\beq
(1 + s_i Q_i) ( 1 + s_j Q_j ) \ge 0,
\eeq
and averaging this, we obtain
\beq
1 + s_i \langle Q_i \rangle   + s_j   \langle Q_j \rangle   + s_i s_j C_{ij}  \ge 0.
\label{LG2}
\eeq
These twelve conditions, which we will call {\it two-time Leggett-Garg inequalities} are necessary conditions for macrorealism at the two-time level. They are also sufficient because if satisfied, the left-hand side of Eq.(\ref{LG2}), multiplied by $\frac{1}{4}$, are precisely the probabilities 
Eq.(\ref{qmom}) matching the given averages and correlation functions.

Finally, the most general possible form of the desired three-time probability is
\bea
p(s_1,s_2,s_3) &=& \frac{1}{8} \left( 1 + s_1 \langle Q_1 \rangle + s_2 \langle Q_2 \rangle +s_3 \langle Q_3 \rangle 
\right.
\nonumber \\
&+&  \left. s_1 s_2 C_{12} + s_2 s_3 C_{23} + s_1 s_3 C_{13} + s_1 s_2 s_3 {D}
\right) .
\label{p123}
\eea
It involves a coefficient $D$, essentially the triple correlator, which is {\it not} measured in the experiment. The question is whether there is any possible value of $D$ for which
\beq
p(s_1,s_2,s_3) \ge 0.
\label{p0}
\eeq
Fine's theorem guarantees that this is indeed possible under the following conditions: the twelve two-time LG inequalities Eq.(\ref{LG2}) hold; the compatibility conditions Eq.(\ref{comp}) hold; and the four three-time LG inequalities, Eqs.(\ref{LG1})-(\ref{LG4}) hold.
The proof of this result is spelled out in detail in Ref.\cite{HalFine}.

The new feature in this protocol, compared to standard LG tests, are the twelve two-time LG inequalities, Eq.(\ref{LG2}). It is these that fill the shortfall in the usual three-time LG inequalities and lead to conditions for MR which are both necessary and sufficient.

Concisely summarized, the protocol just described tests a specific definition of MR, consisting three sets of two-time LG inequalities, one set of three-time inequalities, together with induction and piecewise non-invasive measurability. This is definition of MR is arguably the weakest one possible, and we write,
\beq
\MR_{weak} = \NIM_{pw} \wedge \LG_{12} \wedge  \LG_{23} \wedge \ LG_{13}  \wedge \LG_{123} \wedge \Ind.
\eeq
Like the Bell and CHSH inequalities, it may be satisfied in the face of non-zero interferences, as long as they are not too large \cite{Hal0}.

The twelve two-time and four three-time LG inequalities can be readily simplified by a particular choice of initial state. 
Suppose that we choose the initial state of the system to be at time $t_1$ to be $Q_1 =+1$. Then $C_{12} = \langle Q_2 \rangle $ and $C_{13} = \langle Q_3 \rangle$.
The four three-time LG inequalities Eq.(\ref{LG1})-(\ref{LG4}), then read,
\bea
1 +  \langle Q_2 \rangle + \langle Q_3 \rangle + C_{23} &\ge& 0, 
\label{D1}\\
1 -  \langle Q_2 \rangle - \langle Q_3 \rangle + C_{23} &\ge& 0, \\
1 +  \langle Q_2 \rangle - \langle Q_3 \rangle - C_{23} &\ge& 0, \\
1 -  \langle Q_2 \rangle + \langle Q_3 \rangle - C_{23} &\ge& 0,
\label{D4}
\eea
which therefore coincide with four of the twelve two-time LG inequalities. The remaining eight two-time LG inequalities
consist of trivially satisfied conditions of the form $| \langle Q_i \rangle | \le 1 $. Hence in this simplied situation the four inequalities Eq.(\ref{D1})-(\ref{D4}) are necessary and sufficient conditions for $\MR_{weak}$. Inequalities of this general form have been tested experimentally \cite{EmINRM,SimpLG,Ema}.

The above protocol is readily extended to the four-time situation, for which we find,
\beq
\MR_{weak} = \NIM_{pw} \wedge \LG_{12} \wedge  \LG_{23} \wedge \ LG_{34} \wedge LG_{14}  \wedge \LG_{1234} \wedge \Ind.
\eeq
That is, there are four two-time LG inequalities together with the eight four-time LG inequalities, which have the form
\beq
-2 \le C_{12} + C_{23} + C_{34} - C_{14} \le 2,
\label{LG4*}
\eeq
plus the three more pairs of inequalities obtained by moving the minus sign to the other three possible positions.

\section{Necessary and sufficient conditions for macrorealism using no-signaling in time}

We now review very different conditions for macrorealism which make use of no-signaling in time conditions and do not involve the LG inequalities at all. The most comprehensive version of this approach is that of Clemente and Kofler \cite{Cle} which is followed here.  (Coherence witness conditions \cite{Rob,Wit,Ema} are simpler examples of this approach and conditions similar to the ones that follow have been given by Maroney and Timpson \cite{MaTi}).
They suppose that the system is measured using sequential measurements at three times, with all measurements done in the same experiment and then conditions are imposed to ensure that these measurements are non-invasive, hence we are working with $\NIM_{seq}$. 
This procedure accesses the underlying three-time probability $p(s_1,s_2,s_3)$ directly but the nature of the measurements means that the procedure works only under conditions
considerably stricter than those required in the augmented LG tests.

In the face of potentially invasive sequential measurements, the most general possible form for the three-time probability is
\bea
p_{123} (s_1,s_2,s_3) &=& \frac{1}{8} \left( 1 + s_1 \langle Q_1 \rangle + s_2 \langle Q_2^{(1)} \rangle +s_3 \langle Q_3^{(12)} \rangle 
\right.
\nonumber \\
&+&  \left. s_1 s_2 C_{12} + s_2 s_3 C_{23}^{(1)}  + s_1 s_3 C_{13}^{(2)} + s_1 s_2 s_3 {D}
\right). 
\label{pm}
\eea
Here, the superscripts again acknowledge that the values of averages and correlation functions can depend on whether earlier or intermediate measurements are made. So for example, 
$ \langle Q_3^{(12)} \rangle  $ can depend on whether measurements were made at both $t_1$ and $t_2$, and $C_{13}^{(2)} $ can depend on whether a measurement is made at the intermediate time $t_2$. In contrast to the three-time probability discussed in the LG framework, Eq.(\ref{p123}), here, the triple correlator $D$ is determined by the measurement process.
We assume induction throughout so there is no possibility of dependence on later measurements.

Eq.(\ref{pm}) is non-negative by definition since it is a measurement probability. However, because of the possible dependencies of its components on the context of the measurement, it is not the probability of an independent set of variables, so is not yet the sought after description of macrorealism we seek. Clemente and Kofler therefore imposed a set of NSIT conditions at two and three times to ensure this.

We consider the related measurement probabilities in which measurements are made at only two times, or just one time:
\bea
p_{13} (s_1,s_3) &=&\frac{1}{4} \left( 1 + s_1 \langle Q_1 \rangle +s_3 \langle Q_3^{(1)} \rangle 
+ s_1 s_3 C_{13} \right),
\\
p_{23} (s_2,s_3) &=&\frac{1}{4} \left( 1 + s_2 \langle Q_2 \rangle +s_3 \langle Q_3^{(2)} \rangle 
+ s_2 s_3 C_{23} \right),
\\
p_{12} (s_1,s_2) &=&\frac{1}{4} \left( 1 + s_1 \langle Q_1 \rangle +s_2 \langle Q_2^{(1)} \rangle 
+ s_1 s_2 C_{12} \right),
\\
p_3 (s_2) &=& \half \left( 1 +\langle Q_3 \rangle  \right).
\eea
The NSIT condition
\beq
\sum_{s_2} p_{23} (s_2,s_3) = p_3( s_3),
\label{NSIT23}
\eeq
conveniently denoted $\NSIT_{(2)3}$,
implies that $\langle Q_3^{(2)} \rangle  = \langle Q_3 \rangle $. The NSIT condition
\beq
\sum_{s_1} p_{123} (s_1,s_2,s_3) =p_{23} (s_2,s_3),
\eeq
which we denote $\NSIT_{(1)23}$,
implies that $C_{23}^{(1)} = C_{23}$, $\langle Q_2^{(1)} \rangle =\langle Q_2 \rangle $, and
$\langle Q_3^{(12)} \rangle = \langle Q_3^{(2)} \rangle  $ (which therefore equals
$\langle Q_3 \rangle $).
Finally, the NSIT condition
\beq
\sum_{s_2} p_{123} (s_1,s_2,s_3) = p_{13} (s_1,s_3) 
\eeq
which we denote $\NSIT_{1(2)3}$,
implies $C_{13}^{(2)} = C_{13}$ and $\langle Q_3^{(12)} \rangle = \langle Q_3^{(2)} \rangle  $ (and so they are both equal to $\langle Q_3 \rangle $).
These three NSIT conditions therefore establish that all averages and correlation functions take values independent of whether earlier or intermediate measurements were performed, and the three time probability may then be written:
\bea
p_{123} (s_1,s_2,s_3) &=& \frac{1}{8} \left( 1 + s_1 \langle Q_1 \rangle + s_2 \langle Q_2 \rangle +s_3 \langle Q_3 \rangle 
\right.
\nonumber \\
&+&  \left. s_1 s_2 C_{12} + s_2 s_3 C_{23}  + s_1 s_3 C_{13} + s_1 s_2 s_3 {D}
\right). 
\label{pm2}
\eea
Hence this combination of NSIT conditions are necessary and sufficient conditions for a variety of macrorealism that is clearly stronger than that described in the previous section, and we write,
\beq
\MR_{strong} = \NSIT_{(2)3} \wedge \NSIT_{(1)23} \wedge \NSIT_{1(2)3} \wedge \Ind.
\eeq
From the quantum-mechanical point of view, the NSIT conditions can only hold if the interferences are zero \cite{Hal0}.

In contrast to the LG case, where the measurements are non-invasive by design, the sequential measurements used in these NSIT conditions are invasive in general. In any experimental test it is therefore necessary to adjust the initial state and measurement times (and perhaps other parameters too) to ensure that the NSIT conditions are satisfied. This is why this definition of MR appears to involve far more restrictive conditions than in the augmented LG case, i.e. equalities, rather than inequalities \cite{Cle}. The NSIT conditions are primarily measures of NIM for sequential measurements, whereas NIM is already taken to be satisfied, by design, in the augmented LG case.
Of course, the values of the averages and correlation functions in Eq.(\ref{pm2}) must be the same as those determined in the augmented LG protocol, in Eq.(\ref{p123}), but the conditions under which they can be determined are different in each case: in Eq.(\ref{pm2}) they can be determined only if the equalities consisting of the NSIT conditions hold, whereas in Eq.(\ref{p123}), no such restrictions are required.

The two different types of protocols described in this and the last section are not the only possibilities and clearly delineate the two extremes. A third, intermediate option naturally arises, which is to do three experiments with sequential measurements made at only two times in each case, and then require 
that the three measured two-time probabilities all satisfy two-time NSIT conditions, of the form, Eq.(\ref{NSIT23}); in addition, we then require that the three-time LG inequalities are satisfied. This therefore tests the following version of MR:
\beq
\MR_{int} = \NSIT_{(1)2} \wedge \NSIT_{(1)3} \wedge \NSIT_{(2)3} \wedge \LG_{123} \wedge \Ind
\eeq
Like the augmented LG protocol, it stays close to the spirit of the original LG framework and clearly supplies necessary and sufficient conditions for macrorealism. It requires zero coherence at the two-time level but allows non-zero coherences at the three-time level, as long as they are suitably bounded. This protocol readily extends to the four times, analagous to the augmented LG case for four times.

All the two-time NSIT conditions can be satisfied quite easily, in a quantum-mechanical description,
by choosing an initial state such at $\langle \hat Q(t) \rangle = 0$ at all three times. Also, 
$ \NSIT_{(1)2} $ and $\NSIT_{(1)3} $ can be satisfied by choosing an initial state at $t_1$ diagonal in $\hat Q_1$.
Furthermore, in practice, the NSIT condtitions in $\MR_{int}$ will only be satisfied approximately and it is then necessary to develop extended forms of the LG inequalities appropriate to the case in which there is some signaling. This extension has been carried out by Dzhafarov and Kujala \cite{DzKu} (and is also briefly reviewed in Ref.\cite{HalLG2}).

Clemente and Kofler gave a simple general argument 
that NIM implies MRps \cite{Cle} and the detailed conditions reviewed in this section exemplify this.
(Although note that this argument needs to be adjoined with the assumption that the measurements are  repeatable \cite{Ma3}). However, in view of the different types of NIM characterized in this paper, it is clear that this argument can only involve $\NIM_{seq}$ and not $\NIM_{pw}$.
Their argument refers to measurements made in a single experiment, and not to combinations of results obtained from different experiments. Hence in the LG tests involved in $\MR_{weak}$,  MRps can be violated when $\NIM_{pw}$ holds.

\section{NSIT vs two-time LG inequalities in a quantum-mechanical description}

In a quantum-mechanical description, a direct comparison may be made between the NSIT conditions and LG inequalities for two-times using explicit measurement formulae.
The probability for two sequential projective measurements at times $t_1, t_2$ is,
\beq
p(s_1, s_2) = {\rm Tr} \left( P_{s_2} (t_2) P_{s_1} (t_1) \rho P_{s_1} (t_1) \right),
\label{2time}
\eeq
where the projection operators $P_s (t) $ are defined by $P_s (t) = e^{iHt} P_s  e^{ -i H t } $ and
\beq
P_s = \half \left( 1 + s \hat Q \right).
\eeq
By contrast, the two-time probabilities Eq.(\ref{qmom}) correspond in quantum mechanics to the quantities,
\beq
q(s_1, s_2) = \half {\rm Tr} \left( \left( P_{s_2} (t_2) P_{s_1} (t_1) +  P_{s_1} (t_1) P_{s_2} (t_2) \right) \rho  \right),
\label{quasi}
\eeq
and the two-time LG inequalities Eq.(\ref{LG2}) are simply then $q(s_1, s_2) \ge 0$.
Eq.(\ref{quasi}) may be measured either by measuring the averages and correlation function in various runs, as described, or, as argued in Ref.\cite{HalQ}, more directly, using sequential measurements in which the first one is a weak measurement.
Eqs.(\ref{2time}) and (\ref{quasi}) have the same correlation function and same $\langle \hat Q_1 \rangle$, but differ in the average of $\hat Q (t)$ at the second time.
The sequential measurement probability Eq.(\ref{2time}) does not satisfy the NSIT conditions Eq(\ref{NSIT}) in general. By contrast, Eq.(\ref{quasi}) formally satisfies NSIT, but can be negative.

The relation between these two measurement formulae is given by
\beq
p(s_1,s_2) = q(s_1,s_2) 
+ \frac{1}{8}  \langle [ \hat Q(t_1), \hat Q(t_2)] \hat Q(t_1) \rangle s_2 .
\eeq
The extra term on the right-hand side, which vanishes for commuting measurements, represents interferences (as shown more explicitly in Ref.\cite{HalQ}). If we impose NSIT on $p(s_1,s_2)$ this clearly implies that the interference term is zero and hence that $p(s_1,s_2) = q(s_1,s_2)$. This also means that $q(s_1,s_2) \ge 0 $, which is equivalent to the two-time LG inequalities, Eq.(\ref{LG2}).

However, the converse is not true: $q(s_1,s_2) \ge 0 $ clearly does not imply NSIT for $p(s_1,s_2)$.
Furthermore,
since $p(s_1,s_2)$ is always non-negative, the two-time LG inequalities $ q(s_1, s_2) \ge 0 $ will be satisfied if the interference term is bounded:
\beq
\frac{1}{8}  \left| \langle [ \hat Q(t_1), \hat Q(t_2)] \hat Q(t_1) \rangle \right|   \le p(s_1,s_2)
\eeq
This confirms in this case the general story mentioned earlier (and described at greater length in Ref.\cite{Hal0}): NSIT conditions require zero interference but the LG inequalities, like the CHSH case, require only bounded interference. 

NSIT for $p(s_1,s_2)$ and $q(s_1,s_2) \ge 0 $ are both conditions for MR at two times but they are different types of conditions. NIM is assumed to hold already in the measurement of $q(s_1,s_2)$ and $q(s_1,s_2) \ge 0 $ may therefore be thought as a direct measure of MRps. 
By contrast, NSIT for $p(s_1,s_2)$ measures a combination of NIM and MRps, without being able to distinguish between them.

As an aside, we note the following. Although the role of the quasi-probability $q(s_1,s_2)$ is primarily as a measure of quantumness, when positive, it has the mathematical form of a true probability, and it is therefore natural to ask if it has any interpretation as a probability in the frequentist sense.
The sequentially measured probability $p(s_1,s_2)$ clearly has a frequentist interpretation, but $q(s_1,s_2)$ is determined indirectly. However, it can be argued that when NSIT fails but $q(s_1,s_2) \ge 0 $, then this can be interpreted as a situation in which $p(s_1,s_2)$ is the probability obtained when classically disturbing measurements are used to measure an underlying probability $q(s_1,s_2)$ \cite{HalQP}. Differently put, it is the situation in which the interference effects in sequential measurements which cause NSIT to fail can be modelled as a classical disturbance.

Finally, with the above explicit formulae in hand, we note a connection to coherence witness conditions \cite{Rob,Wit,Ema}.
One can define a witness $W(s_2)$ measuring the degree to which NSIT is violated:
\beq
W(s_2) = \left|  \sum_{s_1} p_{12} (s_1, s_2) - p_2 (s_2) \right|.
\eeq
This is easily seen to be proportional to the interference term,
\beq
W(s_2) = \frac{1}{4}  \left| \langle [ \hat Q(t_1), \hat Q(t_2)] \hat Q(t_1) \rangle \right|.
\eeq
There is therefore a simple relation between the degree of violation of NSIT and the two-time LG inequalities. If the witness is bounded according to
\beq
\half W(s_2)  \le p(s_1,s_2),
\eeq
then $q(s_1, s_2) \ge 0 $. Hence witness conditions, which are usually used to check NSIT, can also be used to check the two-time LG inequalities. This result is also in keeping with the observation in Ref.\cite{WLG} that violations of NSIT have to reach a threshold value before the  LG inequalities are violated.

\section{Summary and conclusions}

The purpose of this paper was to elucidate and compare two very different sets of necessary and sufficient conditions for macrorealism which differ in the way in which they implement the notion of non-invasive measurability. In the first, the weaker form, measurements are made using a number of different experiments, analogous to Bell tests, and then the underlying probability, when it exists, is assembled indirectly. The measurements are non-invasive in a piecewise way, denoted $\NIM_{pw}$.
The probability then exists provided that a set of two and three time LG inequalities  hold (and also induction). This leads to a weak notion of macrorealism, which we write,
\beq
\MR_{weak} = \NIM_{pw} \wedge \LG_{12} \wedge  \LG_{23} \wedge \ LG_{13}  \wedge \LG_{123} \wedge \Ind.
\eeq
In the second, stronger form, proposed by Clemente and Kofler \cite{Cle}, macrorealism 
is defined by a series of NSIT conditions for sequential measurements in which all three measurements are made in the same experiment, together with induction:
\beq
\MR_{strong} = \NSIT_{(2)3} \wedge \NSIT_{(1)23} \wedge \NSIT_{1(2)3} \wedge \Ind.
\eeq
An intermediate notion of MR also naturally arises, in which there are three pairwise experiments with NSIT satisfied for each pair, with all correlation functions required to satisfy the three-time LG inequalities (and induction):
\beq
\MR_{int} = \NSIT_{(1)2} \wedge \NSIT_{(1)3} \wedge \NSIT_{(2)3} \wedge \LG_{123} \wedge \Ind
\eeq
These three conditions have a clear logical connection,
\beq
\MR_{strong} \implies \MR_{int} \implies \MR_{weak},
\eeq
but the converse implications clearly do not hold.
The relation between the NSIT conditions at two times and the two-time LG inequalities was spelled out explicitly in the quantum-mechanical analysis in Section 4. We also noted a relation between the two-time LG inequalities and the degree of violation of coherence witness conditions. 

$\MR_{strong}$ is primarily a measure of non-invasiveness and in the quantum case is satisfied only when the interferences are zero, so is essentially the same type of condition as a number of coherence witness conditions.
By contrast, $\MR_{weak}$ allows non-zero interferences. The measurements are non-invasive by design and hence $\MR_{weak}$ is in effect a direct test of MRps (subject of course to loopholes due to experimental clumsiness \cite{deco}).
Both of these types of macrorealism have been discussed and tested, at least in part, in a number of previous works.
The purpose of the present work has been to make clear that these are different notions of macrorealism, due to the different ways in which NIM is implemented,
although each clearly of interest to explore and test. 

From the perpsective of the consistent histories approach to quantum mechanics 
these different notions of macrorealism correspond to the fact that there exists a hierarchy of classicality conditions. This is utilized and explored in Ref.\cite{HalMC}.

For all of the protocols described in this paper, it would clearly be of interest to check experimentally a full set of necessary and sufficient conditions for macrorealism. This should not be difficult to accomplish with a modest extension of recent experiments: two and three-time LG inequalities have been tested in many different experiments, and likewise two-time NSIT conditions. What is required is an experiment which tests the appropriate combination of such conditions.

\section{Acknowledgements}

I am very grateful Clive Emary, George Knee, Johannes Kofler, Owen Maroney and James Yearsley for many useful discussions and email exchanges about the Leggett-Garg inequalities. I am particularly grateful as always to Thomas Elze for putting together such an excellent conference.

\section{References}


\bibliography{apssamp}

\end{document}